\documentclass[pra, showpacs, floatfix,twocolumn]{revtex4}
\usepackage{graphicx}

\usepackage {amssymb} 
\usepackage {amsmath}
\usepackage[amssymb,thickqspace,mediumspace]{SIunits}

\newcommand{\vect}[1]{{\mathrm {\mathbf #1}}} 
\newcommand{\real}[1]{\mathrm{Re} (#1)} 
\newcommand{\imag}[1]{\mathrm{Im} (#1)} 
\newcommand{\tr}[1]{\mathrm{tr}\, #1} 
\newcommand{\diag}[1]{\mathrm {diag}\, #1} 
\newcommand{\curr}[1]{\unit{#1}{\milli\ampere}}
\newcommand{\temp}[1]{\unit{#1}{\celsius}}

\newcommand{\reffig}[1]{Fig.~\ref{#1}}
\newcommand{\refeq}[1]{Eq.~(\ref{#1})}
 \usepackage{ifthen}
 \newcommand{\refeqs}[2]{\ifthenelse{\equal{#2}{}}{\refeq{#1}}{Eqs.~(\ref{#1})-~(\ref{#2})}}

\begin{document}

\title{Polarization properties in the transition from below to above lasing threshold in broad-area
vertical-cavity surface-emitting lasers}

\author{M. Schulz-Ruhtenberg$^1$, I.~V. Babushkin$^2$,
  N.~A. Loiko$^3$, K.~F. Huang$^4$,
 T. Ackemann$^5$}
\address{$^1$Institute for Applied Physics, University of M\"{u}nster,
  Corrensstr. 2-4, 48149 M\"{u}nster, Germany; now at: Fraunhofer Institute for Laser Technology,
Steinbachstr. 15, 52074 Aachen, Germany \\
$^2$Weierstrass Institute for Applied Analysis and Stochastics
Mohrenstr. 39, 10117,  Berlin, Germany, \\
$^3$Institute of Physics, NASB, 68 Nezalezhnasti Ave., 220072 Minsk,
Belarus, \\
$^4$Department of Electrophysics, National Chiao Tung University,
Hsinchu, Taiwan \\
$^5$SUPA and Department of Physics, University of
Strathclyde, Glasgow G4 0NG, Scotland, UK}

\email{babushkin@wias-berlin.de}

\begin{abstract}
  For highly divergent emission of broad-area vertical-cavity
  surface-emitting lasers (VCSELs) a rotation of the polarization
  direction by up to 90 degrees occurs when the pump rate approaches
  the lasing threshold.  Well below threshold the polarization is
  parallel to the direction of the transverse wave vector and is
  determined by the transmissive properties of the Bragg reflectors
  that form the cavity mirrors.  In contrast, near-threshold and
  above-threshold emission is more affected by the reflective
  properties of the reflectors and is predominantly perpendicular to
  the direction of transverse wave vectors. Two qualitatively
  different types of polarization transition are demonstrated: an
  abrupt transition, where the light polarization vanishes at the
  point of the transition, and a smooth one, where it is significantly
  nonzero during the transition.
\end{abstract}


\maketitle

\section{Introduction}

In the last decades vertical cavity surface emitting lasers (VCSELs)
have played an increasing role in scientific research and
applications \cite{wilmsen99}. One of the features of  VCSEL design
is the possibility to obtain large two-dimensional apertures which
are quite homogeneous and have a small polarization anisotropy.

The polarization behavior in such lasers resulting from the
competition of stimulated and spontaneous emission has been a subject
of many investigations \cite{mulet01a, hermier02,
  willemsen01,shelly00,golubev04}. It is known that the polarization
properties of small and medium aperture VCSELs above
\cite{sanmiguel95b,travagnin96, travagnin97a,
sondermann03a,balle99,panajotov00} and below
\cite{willemsen01,shelly00} threshold are determined mainly by the
intracavity anisotropies.  Well below threshold the polarization
degree is reduced dramatically, in VCSELs \cite{willemsen01} as well
as in edge-emitting semiconductor lasers \cite{ptashchenko96}
(which possess much higher intracavity anisotropies).  However,
below threshold the polarization coincides with the at-threshold
one.

With increasing size of the aperture, off-axis emission becomes
important.  Because the cavity resonance is different for different
transverse modes, the ones with the best alignment between the
cavity resonance and the gain maximum of the active medium
 have the largest gain \cite{moloney90a,jakobsen92}. Recently it was
shown \cite{hegarty99,loiko01,babushkin08} that for strongly
off-axis emission the intra-cavity anisotropies play only an
auxiliary role in polarization selection above threshold. In
contrast, the polarization-selective properties of reflection and
transmission of the distributed Bragg reflectors (DBRs) forming the
cavity mirrors are much more important. Above threshold, the DBR TE
modes (which are in paraxial approximation perpendicular to the
transverse component of the wave vector, i.e.\ s-waves) have higher
\textit{reflectivity}, and the cavity quality factor for the TE
modes is larger than for the TM modes. This determines the
polarization of the above threshold emission, which has an overall
tendency to be perpendicular to the transverse wave vector
(``90-degree rule'') \cite{loiko01}. Later investigations
established that the above-threshold polarization is also strongly
affected by the transverse cavity boundaries, i.e.\ the waveguide
formed by the oxide confinement \cite{babushkin08}. This leads to
strong deviations of the above-threshold polarization state from the
``90-degree rule'' for transverse wave vectors with directions not
parallel to either of the device boundaries \cite{babushkin08}.
The data in \cite{babushkin08} (for lasers with a square
aperture) and in \cite{schulz-ruhtenberg09a} (for lasers with a
circular aperture) indicate also that the polarization direction for
off-axis light is different below and above threshold but there is
no detailed investigation.

In this work we characterize the polarization properties of off-axis
below-threshold emission and show that the polarization direction is
governed mainly by the \textit{transmissive} properties of the top
DBR.  The transmissivity is larger for the TM
Bragg modes (parallel to the transverse wave vector, p-waves) than for
the s-waves, resulting in a ``0-degree rule'' for
polarization selection, i.e.\ the polarization is parallel to the
transverse wave vector.

We consider, both theoretically and experimentally, the transition
from the below-threshold to the above-threshold polarization state
for highly divergent VCSEL emission. The nature of the transition
depends critically on the orientation of the polarization of the
final (lasing) state. When the final polarization obeys the
``90-degree rule'' and is perpendicular to the polarization state
well below threshold, the transition is very abrupt and the light is
unpolarized at the point of transition. On the other hand, when the
polarization of final state is not orthogonal to the initial one,
the transition is considerably smoother, and the emission retains a
relatively large degree of polarization during the transition. Our
theory predicts also that far below threshold the main principal
axes of the intra-cavity and extra-cavity field are perpendicular to
each other due to the strong anisotropic filtering of the light
coupled out via the DBRs.

In the next section we describe the experimental setup; in
Sect.~\ref{sec:exp} the experimental results are reported. In
Sect.~\ref{sec:theory} a theoretical model for the description of the
transition is developed, analyzed and the results compared to the
experimental observations. Concluding remarks are in
Sect.~\ref{sec:concl}.


\section{\label{sec:exp_setup} Experimental setup and methods}
The VCSELs under study are oxide-confined top-emitters with a square
aperture of $\unit{40\times 40}{\micro\meter\squared}$ that are
packaged in TO-type housings without caps. The emission wavelength
is around \unit{780}{\nano\meter}. The lasers consist of two highly
reflective DBRs (top mirror: 31 layers, bottom mirror: 47 layers)
with three \unit{8}{\nano\meter} thick Al$_{0.11}$Ga$_{0.89}$As
quantum wells in between. Together with several
Al$_{0.36}$Ga$_{0.64}$As spacer layers and the GaAs substrate the
whole structure is about \unit{10}{\micro\meter} long. In order to
reduce the electrical resistance of the lasers the interfaces
between different semiconductor layers are graded. A laterally
oxidized layer above the active region provides current and optical
confinement.

The devices are electrically pumped with a low-noise DC current
source between 0 and \curr{30}. The typical lasing threshold at
\temp{0} heat sink temperature is \unit{15}{\milli\ampere}.  The
VCSEL is mounted on a copper plate that is attached to a
thermo-electric cooling element enabling temperature control between
\temp{40} and \temp{-35}. We remark that the actual device
temperature is strongly influenced by the driving current due to
Joule heating effects. Temperature values given here are for the
heat sink only. The device temperature can be inferred indirectly
from optical spectra and changes with current by a factor of about
0.9 K/mA. The device temperature changes the detuning between the
cavity resonance and the maximum gain frequency, since these shift
with distinctly different rates with temperature (typical values are
\unit{0.28}{\nano\meter\per\kelvin} for the gain peak and
\unit{0.075}{\nano\meter\per\kelvin} for the resonance, e.g.\
\cite{hegarty99a}). As considered in detail in
\cite{schulz-ruhtenberg05} this mechanism controls the length scales
of the transverse patterns emitted by the VCSEL.

\begin{figure}[htbp]
  \centering\includegraphics[width=1\columnwidth]{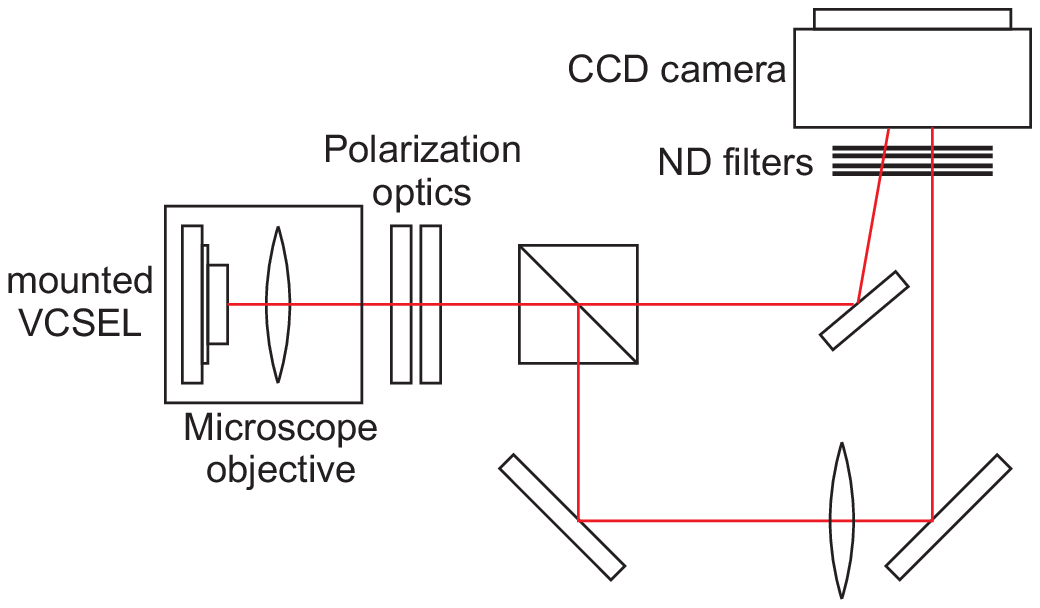}
  \caption{(Color online) Experimental setup. The VCSEL is set into an
    air-tight box to avoid condensation water. Polarization optics:
    half-wave plate and linear polarizer. \label{setup}}
\end{figure}

Figure \ref{setup} shows the setup used for the experiments. The
laser beam is collimated by a microscope objective with a numerical
aperture of 0.8. VCSEL, cooling elements, and the objective are put
into an air-tight box to avoid condensation of water at low
temperatures. The far-field of the laser emission is imaged onto a
high-resolution 14-bit camera with a large charge-coupled device
(CCD) chip.  A half-wave plate and a linear polarizer are inserted
into the beam path. The orientation of the polarizer defines the
reference coordinate system by which the state of polarization is
represented. Horizontal polarization is defined as 0$^\circ$, angles
are measured in counter-clockwise direction. The polarization is
measured by taking far-field images for three settings of the
polarization optics: horizontal ($I_x$), vertical ($I_y$), and
diagonal orientation ($I_{45}$). For the circular component
($I_{circ}$) a quarter-wave plate (set to 45$^\circ$ with respect to
the horizontal) is necessary. From this data, the spatial-resolved
Stokes parameters are calculated:
\begin{equation}
\label{eq:si}
\begin{split}
  & S_0=I_x+I_y,\; S_1=\frac{(I_x-I_y)}{S_0},\; \\ & S_2=\left(\frac{2 \cdot
      I_{45}}{S_0}\right)-1,\; S_3=\left(\frac{2 \cdot
      I_{circ}}{S_0}\right)-1,
\end{split}
\end{equation}
where $S_0$ represents the total  intensity, $S_1$ the (normalized)
amount of light polarized in $x$ (positive $S_1$) respectively $y$
(negative $S_1$) direction, $S_2$ the (normalized) amount of light
polarized along the diagonal direction (positive for 45$^\circ$,
negative for -45$^\circ$) and $S_3$ the (normalized) amount of
circularly polarized light (the sign denotes the direction of
rotation). Using this set of Stokes parameters the degree of
polarization (fractional
  polarization) $p$ and the polarization direction $\varphi$ can be
calculated:
\begin{equation}
  \label{eq:p}
  p=\sqrt{S_1^2+S_2^2+S_3^2}, \,
  \varphi=\frac{1}{2} \cdot \arctan(\frac{S_2}{S_1}).
\end{equation}
The fraction of circular polarization $S_3$ was found to be
of the order of 0.02, thus we assume linear polarization in the
following. In this article we focus on the characteristics of two
VCSELs, which illustrate the general behavior found in the experiments
very well.

\begin{figure}[htbp]
\centering\includegraphics[width=\columnwidth]{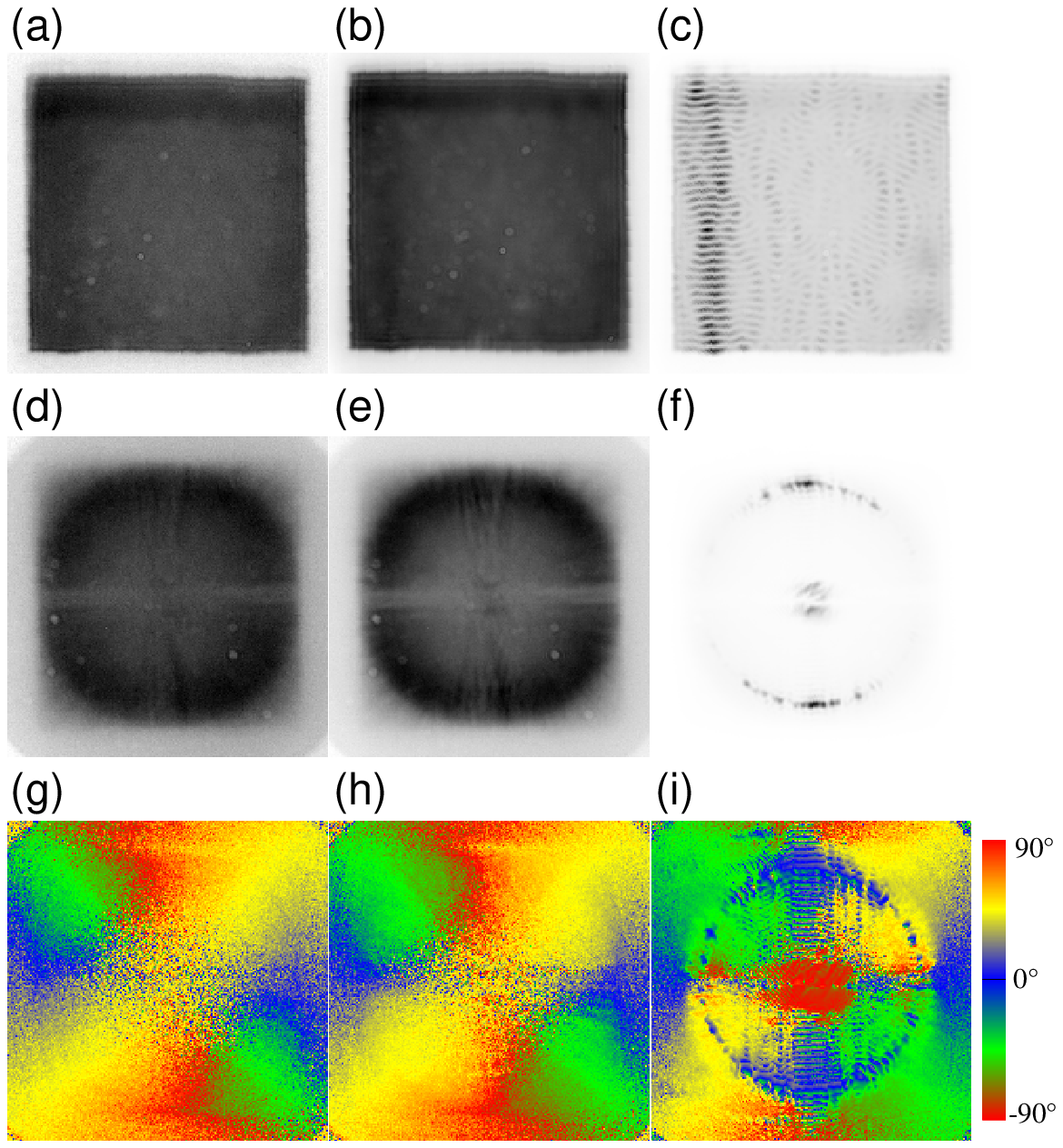}
\caption{ (Color online) Near- and far-field images for device \#1
  below and above threshold. Heat sink temperature $T$=\temp{0} and
  driving current \curr{12.0} (a,d,g), \curr{13.5} (b,e,h), and
  \curr{15.6} (c,f,i). The color-code of the polarization direction
  distribution (lowest row) is shown by the bar on the lower
  right. These images correspond to the transition shown in
  \reffig{trans_plots_40-9}.\label{trans_pics_40-9}}
\end{figure}

\begin{figure}[htbp]
\centering\includegraphics[width=\columnwidth]{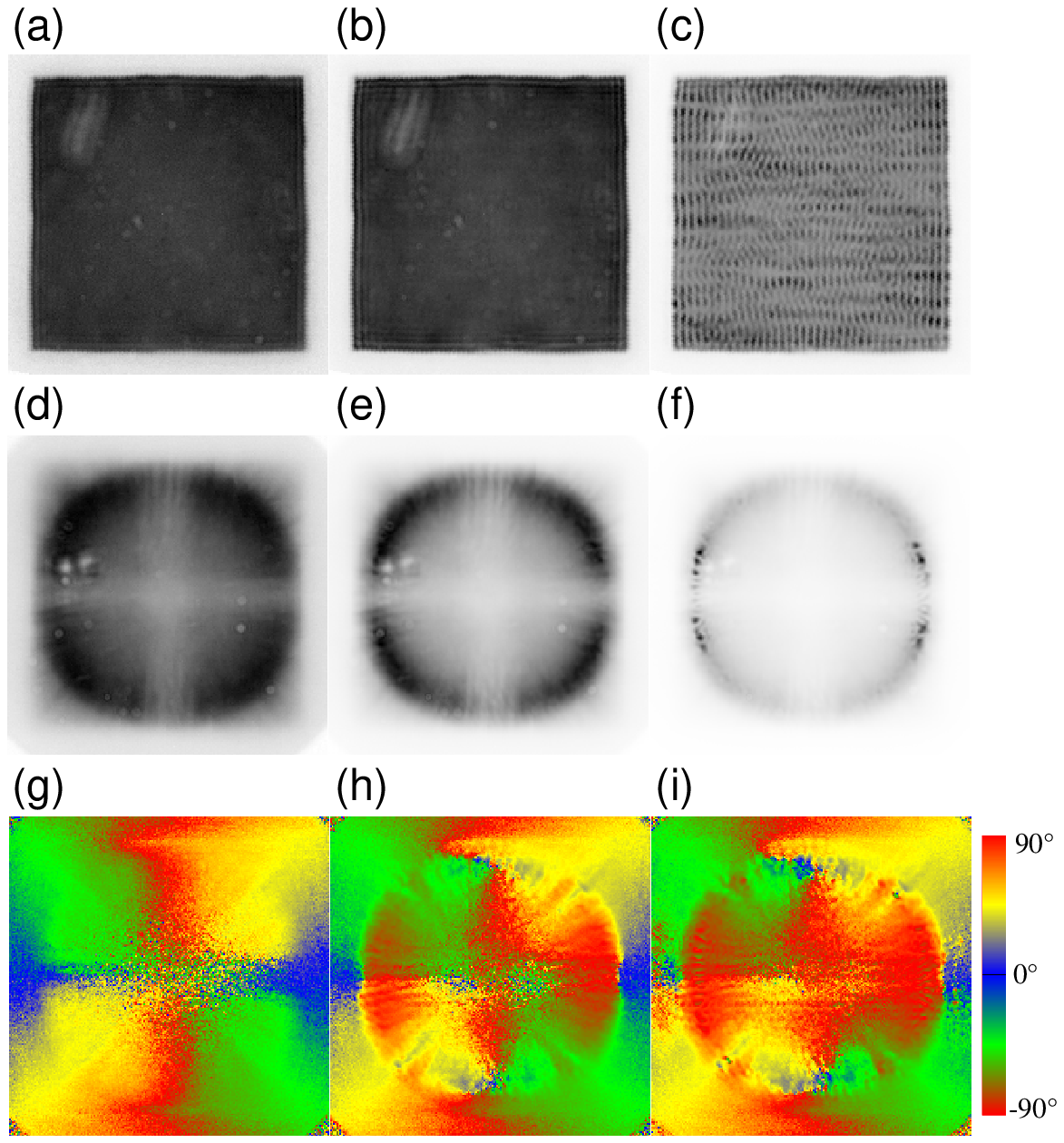}
\caption{(Color online) Near- and far-field images for device \#2 below and above
  threshold. Heat sink temperature $T$=\temp{0} and driving current
  \curr{13.5} (a,d,g), \curr{15.5} (b,e,h), and \curr{16.0} (c,f,i).
  The white spots in (a,b) result from debris on the neutral density
  filters. The color code is the same as in
    \reffig{trans_pics_40-9}.  These images correspond to the
  transition shown in
  \reffig{trans_plots_40-5}.\label{trans_pics_40-5} 
}
\end{figure}

\section{\label{sec:exp} Distribution of spontaneous emission and the
  transition through threshold}
We consider  two nominally identical devices from the design and
growth process, which show however rather different behavior. The
near-field and far-field intensity distributions of the emission are
shown in the figures \ref{trans_pics_40-9} (device \#1), resp.
\ref{trans_pics_40-5} (device \#2). In both cases the heat sink
temperature is $T$=\temp{0}, the threshold current for device \#1 is
\curr{15.2}, for device \#2 \curr{15.6}. The first and second rows
of each figure depict the intensity distributions in grey-scale
coding (black denoting the maximum intensity) of the near-field and
far-field, respectively, the third shows the spatial frequency
distribution of the polarization direction in a cyclic color-code
(red denotes $\pm$90$^\circ$, green -45$^\circ$, blue 0$^\circ$, and
yellow +45$^\circ$, cf.\ to the color bar on the right of
\reffig{trans_pics_40-9}). The columns show the emission below
threshold (panels a, d, g; \curr{12.0} for device \#1, \curr{13.5}
for device \#2), slightly below threshold (panels b, e, h;
\curr{13.5} and \curr{15.5}), and just above threshold (panels c, f,
i; \curr{15.6} and \curr{16.0}). The optical axis is positioned in
the center of each image.

Below and above threshold, the emission has its maximum at a
well-defined wave number. This critical wave number is favored
because it has the most favorable detuning properties as discussed
above (see \cite{schulz-ruhtenberg05} for details about the
dependence of the transverse wave numbers of the emission on the
detuning). Even far below threshold the ring indicating the critical
wave number is easily discernible. With current approaching
threshold it narrows until at threshold the lasing modes develop
from this ring. This is easily explained by the increase of the
Finesse of the cavity if threshold is approached.

\begin{figure*}[htbp]
\centering\includegraphics[width=0.75\textwidth]{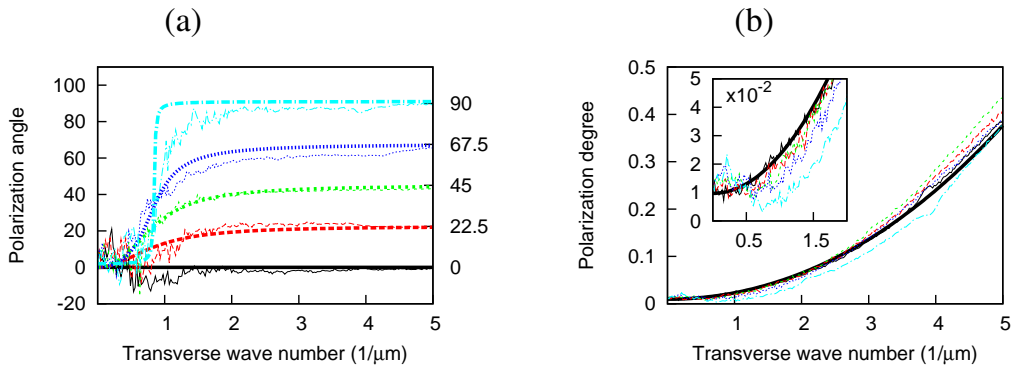}
\caption{(Color online) Radial cuts (at 0$^\circ$ (black solid
line),
  22.5$^\circ$ (red long-dashed line), 45$^\circ$ (green small-dashed line ),
  67.5$^\circ$ (blue dotted line), and 90$^\circ$ (cyan dot-dashed line))
  through the far-field distribution of the spontaneous emission for
  device \#1.  $T$=\temp{0}, $I$=\curr{10}.  (a) Polarization in
  dependence on the transverse wave number $k_\bot$ showing the
  validity of the 0$^\circ$-rule for $k_\bot>2\,\mu$m$^{-1}$.  The thick
  lines indicate the polarization expected from the theory developed
  in Sect.~\ref{sec:theory}.  (b) Fractional polarization in
  dependence on $k_\bot$ for the same cuts as in (a).  The part near
  $k_\bot \approx 0$ is shown in the inset.
    \label{phi_fp_plots}}
\end{figure*}

Within this critical ring, far below threshold the maxima of
spontaneous emission is found at the diagonals in Fourier space, but
moves close to the axes above threshold (i.e.\ with either small
$k_x$ for device \#1 or $k_y$ for device \#2). Just above threshold
VCSEL \#1 emits far-field patterns with two dominant Fourier
components on the y-axis (\reffig{trans_pics_40-9}(f)). The
polarization of these components is in tendency orthogonal to their
wave vector, which is shown in \reffig{trans_pics_40-9}(i), where
the area of lasing emission is polarized horizontally (blue in the
color-code). Below threshold the polarization direction is very
different, parallel to the wave vector (see
\reffig{trans_pics_40-9}(g)). These general observations are also
true for device \#2.

The validity of the `0$^\circ$-rule' is illustrated and tested
further in \reffig{phi_fp_plots}. Here radial cuts through the
polarization distribution at 0$^\circ$, 22.5$^\circ$, 45$^\circ$,
67.5$^\circ$, and 90$^\circ$ with respect to the $x$-axis are shown
for $T$=\temp{0} and $I$=\curr{10} (i.e.\ far below threshold). Each
curve starts at about 10$^\circ$ for $k_\bot\approx 0$ and switches
asymptotically to the angle of the cut. For values of the transverse
wave number above 1~$\mu$m$^{-1}$ the polarization is clearly
parallel to the wavenumber. The deviations amount up to about
10$^\circ$ at 1~$\mu$m$^{-1}$ and decrease for higher wave numbers.
The polarization state at $k_\bot \approx 0$ is interpreted to be
selected by the cavity anisotropies.

It is interesting to note that the curves of the polarization angle
as well as the fractional polarization are continuous till the
maximum value measured of 5~$\mu$m$^{-1}$ though between
3.5~$\mu$m$^{-1}$ and 5~$\mu$m$^{-1}$ (depending on the direction of
the wave vector) the cutoff condition for the transverse modes of
the waveguide formed by the refractive index step (i.e.\ the side
boundaries formed by oxidation) sets in. It is clearly visible in
the center rows of Figs.~\ref{trans_pics_40-9} and
\ref{trans_pics_40-5} that the intensity is cut-off indeed. This
indicates that the influence of the side boundaries on the
polarization characteristics of below-threshold emission is very
small.

The fractional polarization, i.e.\ the amount of linearly polarized
light, in dependence on the transverse wave number is shown in
\reffig{phi_fp_plots}(b), again for the same cuts. It increases
monotonically with wavenumber. The graphs are more or less congruent
which indicates the isotropic character of the phenomenon, showing
its relative independence from the principal axes of the
intra-cavity anisotropy for large enough wavenumbers. The
polarization degree is small, but nonzero for $k_\bot= 0$ (see inset
to \reffig{phi_fp_plots}(b)), which also indicates the influence of
the cavity anisotropies, because the DBRs are isotropic for $k_\bot=
0$.

\begin{figure*}[htbp]
\centering\includegraphics*[width=1.\textwidth]{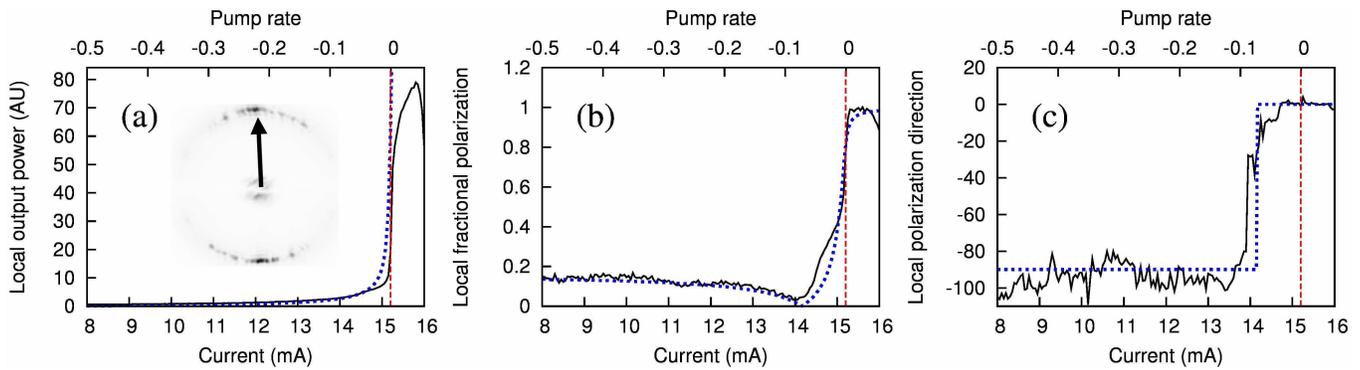}
\caption{(Color online) Transition from spontaneous emission to lasing
  emission of the spot depicted in the inset in (a) for device
  \#1. The black, solid curves are experimental data, the blue,
  dashed ones are calculated from
  Eqs.~(\ref{eq:jls})-(\ref{eq:i}). (a) Local intensity in dependence
  on driving current. The threshold, indicated by the
  vertical red dashed line, is derived from a linear fit to
  the steep slope of the intensity. (b) Degree of polarization. (c)
  Local polarization direction. The change of polarization is evident
  about \curr{1.5} below threshold. \label{trans_plots_40-9}}
\end{figure*}

\begin{figure*}[htbp]
\centering\includegraphics*[width=1.\textwidth]{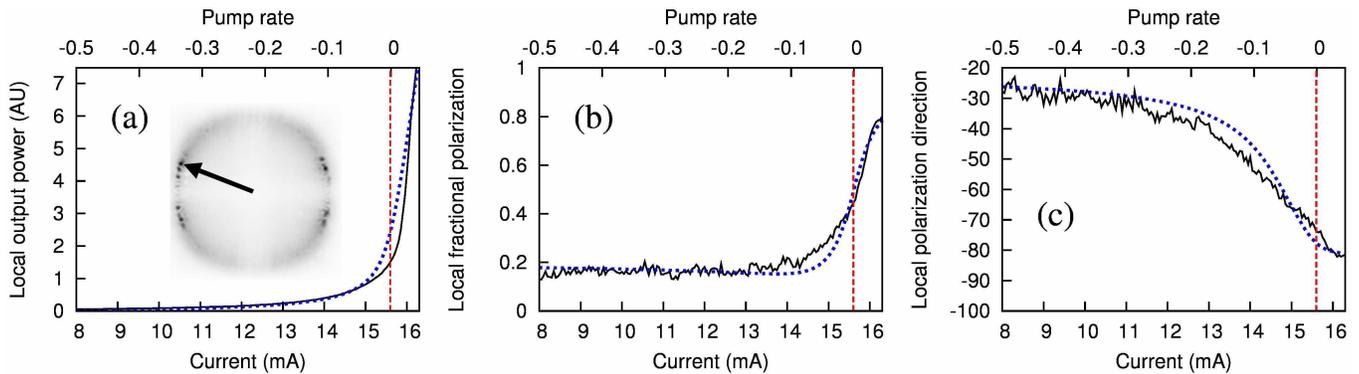}
\caption{(Color online) Transition from spontaneous emission to lasing
  emission of the spot depicted in the inset in (a) for device
  \#2. (a) Local intensity in dependence on driving current. (b)
  Degree of polarization. (c) Local polarization direction. All
  denotations are analogous to
  \reffig{trans_plots_40-9}. \label{trans_plots_40-5}}
\end{figure*}

In the following, we take a closer look at the changes involved in
the transition from spontaneous to lasing emission. This transition
is illustrated in \reffig{trans_plots_40-9} for device \#1 and
\reffig{trans_plots_40-5} for device \#2. Each figure shows in (a)
the local (in Fourier space) intensity, in (b) the fractional
polarization, and in (c) the local polarization orientation in
dependence on the driving current. In addition the pump rate
\begin{equation}
P=\frac{\left(I-I_{th}\right)}{I_{th}}
\end{equation} is displayed on the upper x-axis of the diagrams. The
inset in (a) shows the Fourier component for which these plots were
made (indicated by the arrow). The black lines in the (a) panels
show in both cases the typical behavior of a laser crossing
threshold: A region of low emission intensity representing
spontaneous emission and small slope is followed by a steeply
increasing part indicating lasing emission. Threshold is
extrapolated  by a linear fit to this latter part and indicated by
the dashed red line. The blue curves in \reffig{trans_plots_40-9}
and \ref{trans_plots_40-5} are calculated from equations
(\ref{eq:jls}), (\ref{eq:i}) and will be discussed in the
theoretical section.

The fractional polarization shown in (b) typically follows the
development of the  intensity until it saturates at a maximum of 0.8
to 1.0.

In part (c) of the figures the change of the polarization direction
with current is shown. Far below threshold the polarization angle is
in good agreement with the 0$^\circ$-rule. With increasing current
the scenario is different for the two lasers. For device \#1,  the
polarization starts to change quite abruptly approximately
\curr{1.1} below threshold: In a current range of only \curr{1} the
angle changes to 0$^\circ$, which corresponds to the ``90-degree
rule''. The polarization reaches the target state well below
threshold (about 93\% of $I_{th}$). Note that the fractional
polarization has a pronounced local minimum at the transition.

Device \#2 (\reffig{trans_plots_40-5}(c)) shows a rather gradual
change of the polarization that starts already more than \curr{5}
(at 67 \% of $I_{th}$) below threshold. The behavior is monotonous,
there is no dip in the fractional polarization. This behavior is
typical, if the wavevector is not oriented along one of the two
axis. Only in the latter case (Fig.~\ref{trans_plots_40-9}) an
abrupt transition is found. The continuous transition is  more
typical for the devices under study, though, because the wavevector
configuration depicted in Fig.~\ref{trans_plots_40-5}a is more
typical \cite{babushkin08}.

\section{\label{sec:theory} Theory and discussion}

\subsection{The Ginsburg-Landau equation}

In order to analyze the behavior described above we use a model for a
broad-area VCSEL which accounts for its cavity structure including
Bragg reflectors \cite{loiko01,loiko01b}. For simplicity, we consider
a spatially homogeneous device with an infinite aperture.  For this
case the eigenmodes are plane transverse waves $\vect E_{\vect
  k_t}(x,y,t) = \vect{E(t)} \exp\{i(k_x x+k_y y)\}$ where $\vect
E(x,y,t) \equiv \{E_x,E_y\}$ is the slowly varying complex envelope of
the field inside the cavity, $\vect k_\bot=\{k_x,k_y\}$ is the
transverse component of the wave vector.

Many features of polarization selection at and slightly above
threshold can be obtained by a linear stability analysis
\cite{babushkin04,babushkin08}.  However, for the transition from
below to above threshold it is of critical importance to take into
account both spontaneous emission and the nonlinear saturation.
Therefore we will use here a nonlinear Ginsburg-Landau equation (GLE)
with an additional term describing spontaneous emission (see Appendix
\ref{sec:app} for the derivation). For the spatially homogeneous device with
infinite aperture the equation can be written for every transverse
wave with complex amplitude $\vect E(t)$:
\begin{equation}
  \label{eq:egl0}
  \begin{split}
    \dot{ \vect E} =   - \kappa_{\mathrm{in}} \vect E - \kappa_{\mathrm{out}}
    \Upsilon(\vect k_\bot) \vect E + i \Omega(\vect k_\bot) \vect E +
    \Gamma \vect E - \\  -\kappa_{\mathrm{out}} {\cal G}(\vect k_\bot) I
    \vect E + \vect W.
  \end{split}
\end{equation}
The field decay rate $\kappa =
\kappa_{\mathrm{in}}+\kappa_{\mathrm{out}}$ results from the laser
emission through the DBRs $\kappa_{\mathrm{out}}$ (outcoupling
losses) and intracavity losses $\kappa_{\mathrm{in}}$ by scattering
and absorption. The latter is isotropic (polarization independent),
whereas the former is anisotropic. The anisotropy is described by
the 2$\times$2-matrix $\Upsilon(\vect k_\bot)$, which represents
polarization- and $\vect k_\bot$-dependent losses at the DBRs. In
addition, the matrix $\Upsilon(\vect k_\bot)$ includes
  also the gain in the device (and hence has a component depending linearly on the driving
  current).  $\Omega(\vect k_\bot)$ represents diffraction in the
cavity and in the DBRs, $\cal{G}(\vect k_\bot)$ is a matrix
describing the impact of the nonlinear saturation and $I =
\vect{E}^\dagger \vect E$ is the light intensity ($\dagger$ means
the conjugate transpose). A more detailed description of
$\Upsilon(\vect k_\bot)$, $\Omega(\vect k_\bot)$ and $\cal{G}(\vect
k_\bot)$ is given in Appendix \ref{sec:app}.  $\Gamma$ is the
intracavity anisotropy matrix, which in the basis of the main
anisotropy axes is written as $\Gamma= \diag{(\gamma_a+i \gamma_p,
-\gamma_a-i \gamma_p)}$, where $\gamma_a$ is the amplitude
anisotropy (dichroism), $\gamma_p$ is the phase anisotropy
(birefringence) and $\diag{(\cdot,\cdot)}$ denotes a diagonal matrix
with the corresponding entities on the diagonals.

As indicated above, the other source of anisotropy is the reflection
at the DBRs. The action of the DBRs can be described in terms of s-
and p-waves \cite{born80,babic93}, which are plane transverse waves
with polarization correspondingly perpendicular and parallel to the
direction of the transverse wave vector $\vect k_t$. In this basis,
the matrix of reflection from the $i$th Bragg reflector is diagonal
$R_i(\vect k_t) = \diag{\left\{R_{si}(\vect k_t), R_{pi}(\vect
    k_t)\right\}}$, i.e., pure s- and p- waves are reflected from the
Bragg reflectors without mixing.  The corresponding transmission
matrices $T_i(\vect k_t) = \diag{\left\{T_{si}(\vect k_t),
    T_{pi}(\vect k_t)\right\}}$, $i=1,2$ are also diagonal in this
basis.

The spontaneous emission rate is described by the term $\vect W
\equiv \vect W(\vect k_\bot,t)= \sqrt{\beta_{sp} K j/T_1} \boldsymbol
\xi
(\vect k_\bot,t)$ in \refeq{eq:egl0}. Here $\boldsymbol \xi(\vect
k_\bot,t)$ is a Langevin noise source (see Appendix \ref{sec:app} for
details), $j$ is the normalized current density, $T_1$ is the
population decay time, $\beta_{sp}$ is the spontaneous emission
factor (the fraction of spontaneous emission going into the given
mode), $K$ is the Petermann excess quantum noise factor
\cite{petermann79}, which takes into accounts a possible
non-orthogonality of the modes leading to projection of the noise in
other modes onto the selected one \cite{petermann79}.

\subsection{The coherence matrix}

In  \refeq{eq:egl0} the nonlinear term can be neglected for small
current well below threshold, which results in the linear equation
\begin{equation}
  \label{eq:el0} \dot{\vect E} = - \kappa_{\mathrm{in}} \vect E   -
  \kappa_{\mathrm{out}} \Upsilon (\vect k_\bot) \vect E +
  i \Omega (\vect k_\bot) \vect E + \Gamma \vect E +  \vect W,
\end{equation}
which can be solved directly:
\begin{equation}
  \label{eq:etl}
  \vect E = \exp(\Theta t) \int_0^t \exp(-\Theta \tau) W(\tau) \, d\tau,
\end{equation}
where $\Theta = -\kappa_{\mathrm{in}} - \kappa_{\mathrm{out}} \Upsilon + i \Omega + \Gamma$.

If the field $\vect E$ is known, the Stokes parameters can be obtained
from the coherence matrix $J=\langle \vect E \vect E^\dagger \rangle$
(here $\langle \cdot \rangle$ denotes an ensemble (and not time)
averaging; note also the reverse order of multiplication in this
definition compared to the definition of intensity, which results in a
matrix instead of a scalar):
\begin{equation}
     \label{eq:stokes}
     S_j = \tr( J  \sigma_j),
\end{equation}
where $\sigma_j$ are the Pauli matrices. In particular, the mean
intensity $\langle I \rangle = S_0$ can be obtained as $\langle I
\rangle = \tr J$.

By multiplying \refeq{eq:etl} from the left with its Hermite
conjugate, performing averaging and then integration, we obtain
\begin{equation}
   \label{eq:jl}
   J = - \Theta_r^{-1} J_W  \{1- \exp(\Theta_r t)\} ,
\end{equation}
where $\Theta_r = \Theta + \Theta^\dagger$ and $J_W = \langle \vect W
\vect W^\dagger \rangle$ is the coherence matrix of spontaneous
emission. In the derivation of \refeq{eq:jl} it is taken into account
that the polarization components of $\vect W$ are delta-correlated and
therefore $J_W$ is proportional to the unit matrix $J_{Wij} =
\delta_{ij} K \beta_{sp} j/2 T_1$ (where $\delta_{ij}$,
  $i=1,2$ is a Kronecker delta), i.e., it commutes with all other
matrices. It is also assumed that all functions of matrix arguments
$\Theta$ and $\Theta^\dagger$ commute.

We will search for the statistically stationary solutions of
\refeq{eq:jl} (i.e.\ those with a time-independent coherence
matrix). For such solution being finite, the exponential term in
\refeq{eq:jl} must decay. This is automatically fulfilled below
threshold because $\Theta$ is nothing but a linear stability matrix
for the GLE (\ref{eq:egl0}) without noise far below threshold near its
non-lasing solution, and therefore all the eigenvalues of $\Theta_r$
are less then zero.  Hence, for the small current we obtain:
\begin{equation}
   \label{eq:jls}
   J = - \Theta_r^{-1} J_W.
\end{equation}

\refeq{eq:jls} gives the coherence matrix inside the cavity.  Because
the Bragg reflectors transmit different polarizations differently, the
coherence matrix $J_o$ is different from $J$ for an observer outside
of the cavity:
\begin{equation}
  \label{eq:jo}
  J_o = T J T^\dagger.
\end{equation}

Let us turn our attention to the general case of the nonlinear
\refeq{eq:egl0}. We can simplify the analysis by neglecting the
joint fluctuation of the term $I \vect E$ and replace the intensity
by its mean value $\langle I\rangle$ in this term.  This
approximation can be interpreted in the following way: We replace
the original stochastic process described by \refeq{eq:egl0} by a
simpler, Gaussian one \cite{gardiner04} with the same mean $\langle
\vect E \rangle=0$ and, by its construction, with the same
stationary mean intensity $\langle I\rangle$. We expect therefore
that also the stationary coherence matrix $J$ will not be
significantly altered by this approximation. Of course, the process
described by the original GLE is not, strictly speaking, Gaussian,
especially in the vicinity of threshold or polarization switchings
\cite{mandel95,giacomelli98a}.  However, as we will see later, this
approximation fits rather good to the experimental findings,
allowing at the same time very constructive analytical insight.  By
considering the resulting equation as a linear equation for the
field, and proceeding as above we arrive at \refeq{eq:jl} with the
modified matrix
  \begin{equation}
    \Theta = -\kappa_{\mathrm{in}}- \kappa_{\mathrm{out}} \Upsilon + i \Omega -
    \kappa_{\mathrm{out}} {\cal G} \langle I \rangle + \Gamma.\label{eq:theta}
\end{equation}
As before, we consider only the finite statistically stationary
solutions, for which the exponential term in \refeq{eq:jl} decays, and
obtain again \refeq{eq:jls}, with $\Theta$ given now by
\refeq{eq:theta}.

Taking the trace of \refeq{eq:jls} one obtains an implicit equation
for the intensity $\langle I \rangle$:
 \begin{equation}
   \label{eq:i}
   \begin{split}
     \langle I \rangle = \tr
     \left\{\left(2\kappa_{\mathrm{in}}+\kappa_{\mathrm{out}}
         (\Upsilon+\Upsilon^\dagger) - (\Gamma+\Gamma^\dagger) +
       \right. \right. \\ \left. \left. + \kappa_{\mathrm{out}}
         \langle I \rangle ({\cal G}+{\cal G}^\dagger)\right)^{-1} J_W
     \right\},
   \end{split}
\end{equation}
which is an equation of third order for $\langle I \rangle$. The
intensity outside the cavity is then given by $\langle I_o
\rangle=\tr{\{T J T^\dagger \} }$. \refeq{eq:i} has only one positive
root, which is small ($\langle I \rangle \approx 0$) below threshold
and grows asymptotically linearly with current ($\langle I \rangle
\sim j$) above threshold.  Below and at threshold it is however in
rather good agreement with experiment (compare the black and blue
curves in \reffig{trans_plots_40-9}(a) and
\reffig{trans_plots_40-5}(a)).

\subsection{General features}

As stated before, the most significant difference between the
devices \#1 and \#2 is in the pattern formed above threshold.
Whereas the below-threshold state obeys the ``0-degree rule'' for
both devices (i.e.\ the polarization direction is parallel to $\vect
k_\bot$), the polarization of the above threshold state deviates
significantly from the ``90-degree rule'' for device \#2 (around
30$^\circ$).  In \cite{babushkin08} it was shown that for high
enough $k_\bot$ the polarization direction is aligned mainly to the
transverse side boundaries of the device rather than to the
anisotropies (either intra-cavity or Bragg-induced ones). The
validity of the ``90-degree rule'' above threshold depends therefore
on the position of the spots in the far field with respect to the
directions of the side boundaries. For a device with the boundaries
parallel to the $x$ and $y$ coordinate axes the ``90-degree rule''
is satisfied when the spots are located close to the $x$ or $y$
axes, whereas for spots away from the axes the polarization
direction deviates from the ``90-degree rule''. Note that for the
purpose of this paper it is not important why some devices emit in a
specific wave vector configuration and some in another. We are only
exploring the consequence of a given wave vector configuration on
the polarization behavior.

\begin{figure}[htbp]
  \centering\includegraphics[width=0.35\textwidth,clip=]{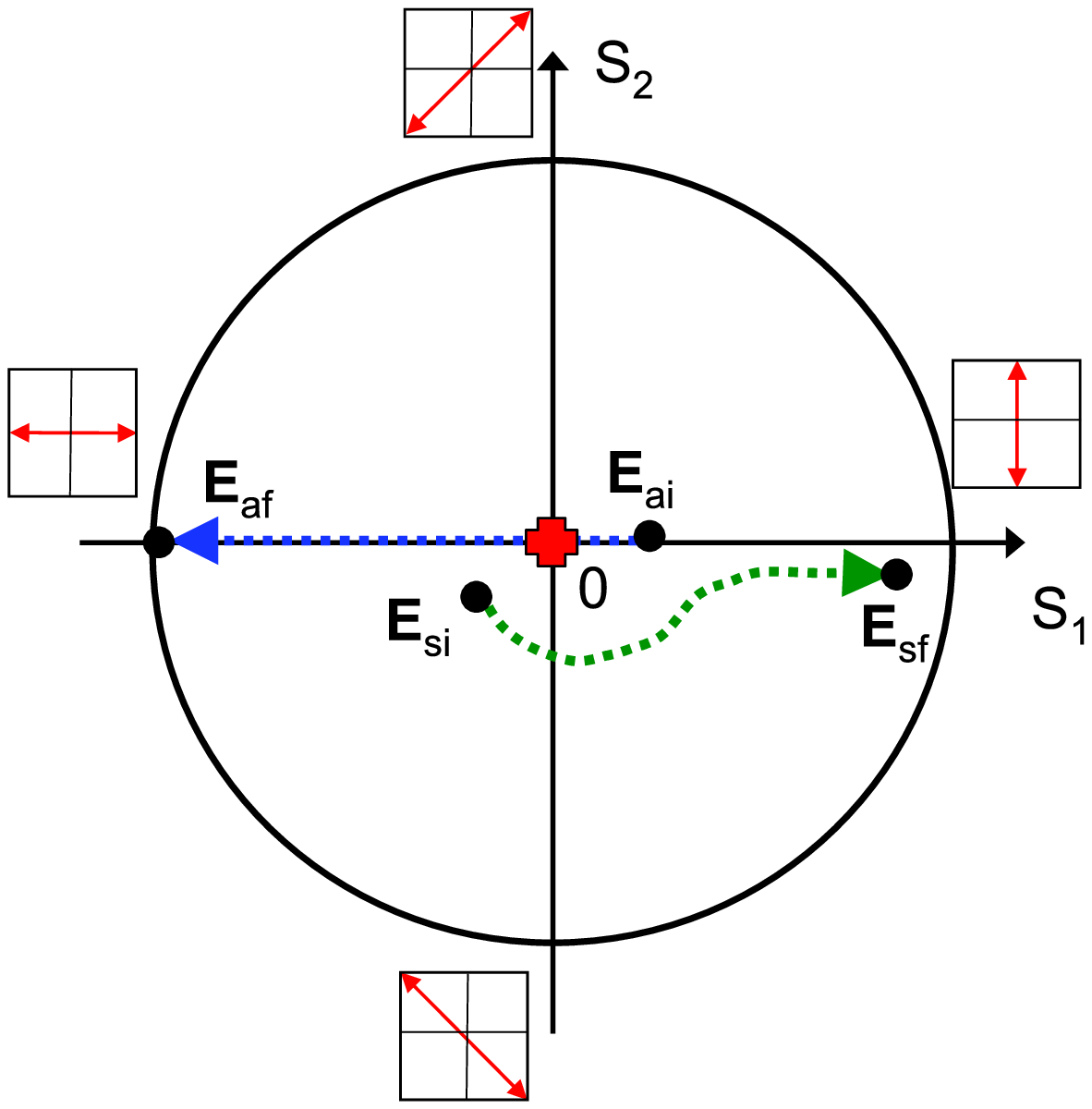}
  \caption{\label{bloch} (Color online) Different types of
  transitions
    from non-lasing to lasing state in the space of Stokes parameters
    ($S_1$, $S_2$): Abrupt transition (blue straight dashed
      arrow from $\vect E_{ai}$ to $\vect E_{af}$, corresponds to
      device \#1) and smooth transition (green curved dashed arrow
      from $\vect E_{si}$ to $\vect E_{sf}$, corresponds to device
      \#2).  The initial polarization state for the abrupt transition
    $\vect E_{ai}$ is orthogonally polarized to the final one $\vect
    E_{af}$, whereas for the initial $\vect E_{si}$ and final $\vect
    E_{sf}$ states for the smooth transition this is not the case.
    For convenience, the polarizations directions for different
    quadrants of ($S_1$, $S_2$)-plane are shown by red arrows in the
    black boxes. The red cross in the center corresponds to a fully
    unpolarized state ($p=0$) whereas the black circle shows
    the fully polarized state ($p=1$).}
\end{figure}

The difference between these two situations can be depicted in the
space of Stokes parameters (\reffig{bloch}).  Because the
polarization is always linear, $S_3=0$ and we use only its
two-dimensional subspace ($S_1$, $S_2$).  One can see that -- if the
initial state $\vect E_{ai}$ and the final state $\vect E_{af}$ are
orthogonal to each other-- $S_2$ experiences a zero crossing during
the transition (red cross in \reffig{bloch}). According to
\refeq{eq:p} this means that the degree of polarization is also zero
at the crossing point. The polarization direction retains only two
values during this transition. Hence the transition, which appears
at the point $S_1=S_2=0$, is an abrupt switching between two
discrete polarization states. Both the transition of the degree of
polarization through zero and the abrupt switching can be seen in
\reffig{trans_plots_40-9}(b,c).  On the other hand, when the final
state $\vect E_{sf}$ is not orthogonal to the initial one $\vect
E_{si}$, the polarization follows a path avoiding the origin, i.e.,
it rotates to the final state instead of switching to it. In this
case, the degree of polarization does not pass through zero and the
polarization direction changes smoothly [see
\reffig{trans_plots_40-5}(b,c)].

We will refer to the first case as an ``abrupt'' transition, and to
the second case as a ``smooth'' one. One can see a remote analogy to
the Ising and Bloch transitions \cite{coullet90a}, which also
represent abrupt ``switching'' or smooth ``rotating'' behavior.  The
role of the vector of magnetization in this case is played by the
Stokes parameters. One should note however that Ising and Bloch
transitions occur in space, i.e., between energetically equivalent
spatially separated states (at constant external parameters),
whereas here the transition takes place in dependence on an external
parameter (current), i.e. in the parameter space.

The above mentioned difference between the abrupt and smooth
transitions can be expressed more directly in terms of the coherence
matrix. For the abrupt transition the coherence matrix is diagonal in
\textit{one and the same coordinate basis} during the whole
transition. 
Only two polarization directions are possible, depending on which
diagonal element $J_{11}$ or $J_{22}$ is larger. The point
representing the polarization state in Fig.8 can move only along a
straight line passing through zero $(S_1=0,S_2=0)$. At the point of
transition $J_{11}=J_{22}$, i.e. the light is unpolarized, and the
polarization direction changes abruptly.
In the case of a smooth transition the situation is different.  Of
course, the coherence matrix is Hermitian and therefore there is
always a Cartesian coordinate basis in which it is diagonal. However,
this basis \textit{is changing} during the transition.  That reflects
the existence of several competing mechanisms (DBRs, side boundaries,
intra-cavity anisotropies), each of them having its own principal
axes. Their mutual influence is changing with a chance of parameter
(here current). In this case, the Stokes parameters do not vary along
a straight line and can avoid the zero crossing.  It should be noted
also that the origin $(S_1=0,S_2=0)$ is a special point in the sense
that the coherence matrix in this point is proportional to the
identity matrix and therefore is diagonal in every coordinate basis.

\subsection{Detailed discussion}

The polarization, intensity and fractional polarization obtained
from \refeq{eq:jo}, \refeq{eq:jls} and \refeq{eq:i} are shown in
\reffig{phi_fp_plots}, \ref{trans_plots_40-9}, and
\ref{trans_plots_40-5} in comparison to experimental data.  The
parameters of the active layer and the cavity used for calculations
are $\alpha = 3$, $\gamma=10^{3}$ ns$^{-1}$, $\gamma_a=0.8$
ns$^{-1}$, $\gamma_p=0$. The DBRs, consisting of 31 (top mirror) and
47 (bottom mirror) $\lambda/4$ layers of material with alternating
refractive index $n_1=3.46$, $n_2=3.093$  and the transparency
current $J_{tr}=7$ mA were assumed. The effective round-trip time
  $\tau = 33$ fs includes also the effects of the dispersion in the DBRs
  and in the cavity \cite{schulz-ruhtenberg05}. The outcoupling loses
for these parameters are $\kappa_{\mathrm{out}}=17.4$ ns$^{-1}$.  The best
coincidence with the experimental results appears for
$\kappa_{\mathrm{in}}=65$ ns$^{-1}$, which is in rather good agreement
with estimations based on losses in the p-DBR and typical gain values
of GaAs quantum wells \cite{babic97,coldren95}.

In the framework of the theory presented here only an infinite
device can be considered. Therefore, the final lasing state
satisfies always the ``90-degree rule'' \cite{loiko01}. As the laser
approaches threshold, the influence of the side boundaries start to
play an important role \cite{babushkin08}, and the polarization may
not be perpendicular to $\vect k_\bot$ anymore.  For the device \#1
this deviation is very small but it is not negligible anymore for
the device \#2.  We take this into account in our model by
introducing an artificial rotation of the main axes of the matrix
$\Theta$ for the device \#2, so it is becomes diagonal not in the
basis of s- and p-waves but in another, rotated one. On the other
hand, we keep the transition matrix $T$ the same (i.~e. diagonal in
the basis of s- and p-waves). This makes the resulting coherence
matrix non-diagonal. The angle of rotation is 25$^\circ$, which
corresponds approximately to the angle of the wavevectors 
visible in the inset of \reffig{trans_pics_40-5}(a).

Although the spontaneous emission factor is rather small
($\beta_{sp} \approx 10^{-5}$) for VCSELs with a transverse size
around 40 $\mu$m \cite{shin97}, it can be significantly enhanced by
the Petermann excess factor. The value of $K$ depends extremely
strongly on the inhomogeneities and imperfections in the
construction of a particular device. The best results in comparison
to experiment give the values $K \beta_{sp}=6 \times 10^{-4}$ for the
device \#1 and $K \beta_{sp}=10^{-2}$ for \#2. 
In \cite{babushkin08} it is shown that the four spots at the corners
of a rectangular which form the dominant Fourier peaks of the
spatial structures (see inset of \reffig{trans_plots_40-5}(a)) can
form the eigenmode of the transverse waveguide but are not
simultaneously an eigenmode of the reflection operator of the DBR.
Thus the   reflection couples many transverse wavevectors $\vect
k_\bot$. This
  may be the the origin of the rather large Petermann factor $K$,
  especially for device \#2. In addition, non-orthogonality of the modes
  might originate from inhomogeneities of the structure and current
  distribution (which is clearly visible in the intensity
  distributions in \reffig{trans_pics_40-9}(a-c),
  \reffig{trans_pics_40-5}(a-c), for both device \#1 and \#2).

\begin{figure*}
  \centering\includegraphics[width=1.\textwidth]{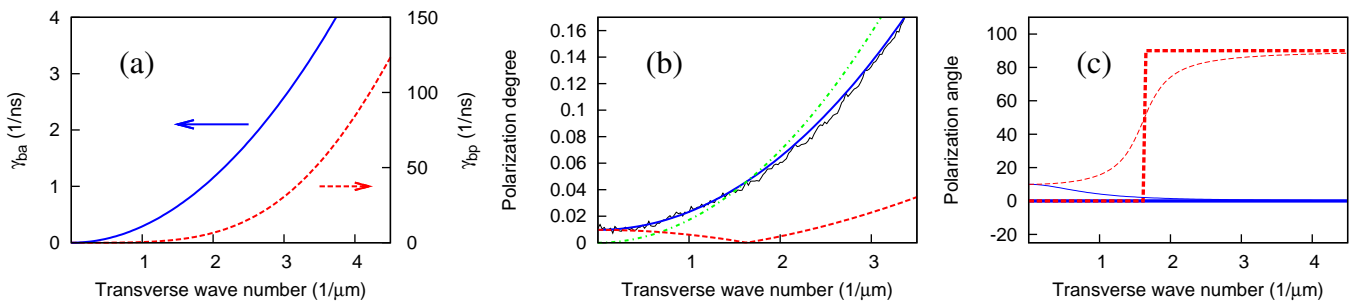}
  \caption{\label{fig:br-anisotr}(Color online) (a) The amplitude
    $\gamma_{ba}$ (blue, solid line) and phase $\gamma_{bp}$ (red,
    dashed line) anisotropy induced by the Bragg reflector. (b) The
    degree of polarization for the parameters of device \#1 (see
    \reffig{phi_fp_plots}) and $k_y=0$ according to theoretical
    calculations. The polarization degree inside the cavity is
    indicated by the red dashed line, outside the cavity by the blue
    solid line. The corresponding experimental curve is depicted by
    the black line. The extra-cavity polarization in the ``pure
    filtering limit'' (i.e.\ assuming completely unpolarized light
    inside the cavity) is indicated by the green dot-dashed line.  (c)
    The polarization direction inside (red dashed lines) and outside
    (blue solid lines) the cavity for the intra-cavity anisotropy
    favoring $x$-polarization (thick lines, as in panel b) or an
    anisotropy favoring a polarization direction of 10$^\circ$ (thin
    lines).  }
\end{figure*}

With this choice of parameters, a very good agreement
between experiment and theory is obtained for the development of the
fractional polarization and the local polarization direction versus
current for both devices (Figs.~\ref{trans_plots_40-9}, and
\ref{trans_plots_40-5}) as well as for the dependence on the
transverse wave vector (Fig.~\ref{phi_fp_plots}).  The data reflect
the degree of abruptness of the transition
(Fig.~\ref{trans_plots_40-9}c vs.\ Fig.~\ref{trans_plots_40-5}c) and
the monotonous vs.\ non-monotonous development of fractional
polarization (Fig.~\ref{trans_plots_40-5}b vs.\
Fig.~\ref{trans_plots_40-9}b). The increase of fractional
polarization and the convergence towards the ``90$^\circ-$rule''
with increasing wave number is due to the increasing anisotropy
between s- and p-waves, of course (see the blue line in
Fig.~\ref{fig:br-anisotr}a).

The case of the abrupt transition allows more analytical insight.
Let us suppose for the sake of clarity that the isotropic
intra-cavity anisotropy is diagonal in the representation of s- and
  p-waves.  In this situation, all the matrices in \refeq{eq:jls} and
\refeq{eq:jo} are diagonal in this representation. Therefore the
coherence matrices $J$, $J_o$ are also always diagonal. In this case
\refeq{eq:jls} and \refeq{eq:jo} are decoupled into independent
equations for the diagonal elements:
\begin{gather}
  J_{ii} = \frac{ K \beta_{sp} j}{2 T_1 \{\kappa_{\mathrm{in}}+
    \kappa_{\mathrm{out}} \real{\Upsilon_{ii}} + \real{\Gamma_{ii}} +
    \kappa_{\mathrm{out}} \langle I \rangle \real{{\cal G}_{ii}}\}},
  \, \notag \\ J_{oii}= |T_{ii}|^2 J_{ii}, \label{eq:jdiag}
\end{gather}
where $i=1$ corresponds to the DBR s-wave whereas $i=2$ to
the p-wave.

Well below threshold the denominator in \refeq{eq:jdiag} is positive
and approximately the same for both s- and p- waves (as it will be
discussed below), and the light outside of the cavity is slightly
p-polarized ($J_{o11} < J_{o22}$) due to the filtering by the Bragg
reflector. As the current approaches threshold the denominator for
the s-wave in \refeq{eq:jdiag} tends to zero whereas the one for the
p-wave remains positive, which provides superiority for the s-wave
strongly overcoming the opposite difference in transmission.
Obviously, at some point the transition between s- and p-polarization
occurs. At the point of transition $J_{o11}=J_{o22}$ (and therefore
$S_1=0$, $S_2=0$), i.e., the output is unpolarized (see
\reffig{bloch}, blue straight arrow, and
\reffig{trans_plots_40-9}(b)). Because the coherence matrix is
diagonal, the polarization direction can have only two values,
corresponding to either $J_{o11}<J_{o22}$ (``0-degree rule'') or
$J_{o11}>J_{o22}$ (``90-degree rule''). Therefore, at the transition
point, the polarization direction changes abruptly and is constant in
all other points above and below threshold (see
\reffig{trans_plots_40-9}(c)).

  Let us now consider the behavior of the extracavity polarization far
  below threshold (when the intensity-dependent term in
  \refeq{eq:jdiag} can be neglected). If we suppose that the
  intra-cavity losses $\kappa_{\mathrm{in}}$ are much larger than the
  ones through outcoupling $\kappa_{\mathrm{out}}$ and than the
  intracavity anisotropy $\Gamma_{ii}$ we obtain:
\begin{equation}
   \label{eq:j0}
   J_{ii} = \frac{ K \beta_{sp} j}{2 T_1 \kappa_{\mathrm{in}}}, \, J_{oii}= |T_{ii}|^2 J_{ii}.
\end{equation}
In this case the output polarization is governed by the transmission
through the DBR $|T_{ii}|^2$, and the laser cavity works as a simple
filter for almost unpolarized intra-cavity radiation.  Because
$|T_{s}|^2>|T_{p}|^2$, this results in the ``0-degree rule'' due to
the dominance of p-polarization direction in transmission.  In this
case (we will call it ``pure filtering case'' here), the degree of
polarization is determined only by the transmitting properties of the
Bragg reflector (see \reffig{fig:br-anisotr}(b), green line).
The difference in transmission between s- and p-waves
  increases with $k_\bot$ approximately quadratically, and the degree
  of polarization in this case reflects this
  dependence according to \refeq{eq:j0}.

 For the opposite case $\kappa_{\mathrm{out}} \gg
\kappa_{\mathrm{in}}$ and assuming a diagonal coherence matrix, we
obtain
\begin{equation}
    J_{oii} \sim
    |T_{ii}|^2/\real{\Upsilon_{ii}},
\end{equation}
instead of \refeq{eq:j0}. Because for small current $\Upsilon_{ii}
\sim 1-|R_{ii}|^2=|T_{ii}|^2$, the light outside the cavity is
completely unpolarized in this case.  This is in agreement with the energy
conservation principle, since the numbers of intracavity photons in
both polarizations are increased by the spontaneous emission equally
and the polarizations do not mix. Therefore, in the stationary case in
the absence of intracavity losses the energy escaping the cavity must
be also equal for both polarizations.

In general, the extracavity polarization degree, defined by
  \refeq{eq:jdiag}, lies in between these two limiting cases
  (\reffig{fig:br-anisotr}(b), thick, solid blue line).  As the Bragg
  reflection is isotropic for $k_\bot=0$
  ($\Upsilon_{ii}\left. \right|_{\vect k_\bot=0}=1$ for $i=1,2$), the
polarization for $\vect k_\bot \approx 0$ is defined only by the
intra-cavity amplitude anisotropy $\gamma_a$ (and does not depend on
$\gamma_p$).  The best agreement with the experimental value of the
degree of polarization for $\vect k_\bot=0$ is achieved for
$\gamma_a=0.8$~ns$^{-1}$, a reasonable number in line with typical
observations in small-area VCSELs \cite{exter98,sondermann04b}. With
increasing transverse wavenumber $k_\bot$, the relative importance
of the anisotropy induced by $\Upsilon(\vect k_\bot)$ increases. The
amplitude anisotropy $\gamma_{ba}(\vect
k_\bot)=\kappa_{\mathrm{out}} (\real{\Upsilon_{11}(\vect
k_\bot)}-\real{\Upsilon_{22}(\vect
  k_\bot)})/2$ and the phase anisotropy $\gamma_{bp}(\vect
k_\bot)=\kappa_{\mathrm{out}} (\imag{\Upsilon_{11}(\vect
  k_\bot)}-\imag{\Upsilon_{22}(\vect k_\bot)})/2$, induced by
$\Upsilon(\vect k_\bot)$ are shown in \reffig{fig:br-anisotr}(a).  In
analogy to the intra-cavity anisotropy, only $\gamma_{ba}$ plays a
role in the determining of the coherence matrix. As one can
  see, $\gamma_{ba}$ increases approximately quadratically with
  $k_\bot$. The degree of polarization \textit{inside} the cavity (see
  \reffig{fig:br-anisotr}(b), thick, dashed red line) is influenced
  more strongly by the intracavity anisotropy for small $k_\bot$
  (below $\sim 1.5$ $\mu$m$^{-1}$) and by the Bragg-induced anisotropy
  for larger $k_\bot$. But it remains relatively small for all
  $k_\bot$, and the polarization degree \textit{outside} of the cavity
  is therfore quite strongly determined by the Bragg filtering
  mechanism (see \reffig{fig:br-anisotr}(b), thick, solid blue line).

  Now, let us consider the behavior of the polarization of
  \textit{intra-cavity} field in dependence on $\vect k_\bot$ far
  below threshold in more detail. As an example, a cut displaying the
  polarization angle along the $k_x$ axis, assuming the anisotropy
  also being directed along the $x$-axis is shown in
  Fig.~\reffig{fig:br-anisotr}(c). In this case, the intracavity and
  extracavity light for $k_\bot \approx 0$ is x-polarized. However,
  for high enough $k_\bot$ (above $\sim 1.5$ $\mu$m$^{-1}$) the
  DBR-induced anisotropy overcomes the intracavity one and the
  intra-cavity light becomes weakly polarized in the direction
  \textit{perpendicular} to $\vect k_\bot$ (see
  \reffig{fig:br-anisotr}(c), thick red, dashed line).  This degree of
  polarization is small and canceled when the light is transmitted
  through the Bragg reflector.  Hence the polarization of the light
  outside of the cavity is still determined by the transmission, i.e.,
  is \textit{parallel} to $\vect k_\bot$, as it mentioned above (see
  \reffig{fig:br-anisotr}(c), blue lines). 

Hence, we also encounter a polarization transition in the
intracavity field, if we consider $k_\bot$ as a parameter instead of
the current. This transition is abrupt (see
\reffig{fig:br-anisotr}(c), thick red, dashed line) and the
fractional polarization vanishes at the transition point (see
\reffig{fig:br-anisotr}(b), thin red, dashed line). It can be
understood in a way fully analogous to   the abrupt transition
appearing with the change of current. It should be noted, that this
transition is not apparent outside the cavity (thick blue line in
\reffig{fig:br-anisotr}(c)).

This behavior becomes slightly more complex, if one analyzes the
polarization along a cut in Fourier space whose direction does not
coincide with the preferred intra-cavity anisotropy axis. In that
case, the change of polarization angle vs.\ wavenumber is smooth
(\reffig{fig:br-anisotr}(c), thin red line) and also detectable
outside the cavity (\reffig{fig:br-anisotr}(c), thin blue line).
This is backed up by the experimental curves displayed in
Fig.~\ref{phi_fp_plots}a, where the transition is most notable for
the cyan curve, which corresponds to a cut along 90$^\circ$; i.e.,
the competition between intra-cavity anisotropy and Bragg anisotropy
is more apparent in the extra-cavity polarization, if the deviation
between wavevector and anisotropy angle increases.

\section{\label{sec:concl} Conclusion}

For highly divergent emission in wide aperture VCSELs the
  polarization direction is determined by the properties of the
  distributed Bragg reflectors (DBRs), and (close to threshold) by the
  transverse structure of the cavity.  Far below threshold the
  polarization is independent from the cavity structure and the
  p-wave of the DBRs (which has the polarization direction
parallel to $\vect k_\bot$) prevails in the emission because the
almost unpolarized intra-cavity light is filtered by the
transmission through the DBR, which is higher for p-waves than for
s-waves. In the simplest
  case, when the above threshold polarization direction is not
  strongly influenced by the transverse boundaries, an abrupt switch
  of polarization direction occurs as the current increases towards
  the laser threshold: the s-wave inside the cavity start to prevail
  because the reflectivity and therefore the quality factor of the
  cavity is higher compared to the one for p-wave. At the point of
  transition the filtering effect of the transmission through the DBR
  can not compensate the intra-cavity polarization anymore, and the
  polarization changes its direction from the one parallel to $\vect
  k_\bot$ to the perpendicular one. This abrupt change of polarization
  direction is accomplished by the passing through zero of the degree
  of polarization of the light outside of the cavity.

On the other hand, if the above-threshold polarization does
  not coincide with the DBR s-mode, the transition is qualitatively
  different. In this case, the polarization changes smoothly and the
  degree of polarization is significantly nonzero during all the
  transition. As was shown in \cite{babushkin08}, the deviation of the
  above-threshold polarization direction from the one dictated by the
  s-wave of the DBRs can be due to the influence of the side
  boundaries of the cavity.

The difference between two types of transition can be
  explained in the terms of coherence matrix. The abrupt transition
  occurs when the coherence matrix of the light outside the cavity is
  diagonal (in one and the same basis) during the whole transition. In
  this case, only two polarization directions are possible. If the
  different mechanisms influencing the polarization have different
  directions, the resulting coherence matrix is non-diagonal and
  arbitrary polarization direction is possible.

  We note an analogy between the abrupt and smooth transitions
  described here to Ising and Bloch transitions between two equivalent
  states in ferromagnetics with respect to a ``switching'' vs.
  ``rotation'' behavior.  However, the polarization transition in this
  article occurs in parameter space (between energetically
  inequivalent states) in contrast to the usual Bloch and Ising ones.

\section*{Acknowledgements:} We acknowledge financial support from the
Deutsche Forschungsgemeinschaft, the Deutsche Akademische
Austauschdienst and the Taiwanese Research Council (grant
NSC-96-2112-M-009-027-MY3) at the initial stages of the work.

\appendix

\section{\label{sec:app} The Ginsburg-Landau equation}

\subsection{The initial equations}

Polarization phenomena in VCSEL are often modeled using an equations
for the intracavity field $\vect E$ and the total carrier density $D$
as well as the population difference $d$ between sub-bands with
opposite carrier spin \cite{sanmiguel95b,loiko01}.  We start from the
nonlinear equations (18) of \cite{loiko01} for the normalized
complex-valued envelope $\vect E(t, \vect k_\bot)$, of the optical
field for given $\vect k_\bot$ and carrier population variables $D$,
$d$:
\begin{widetext}
 \begin{eqnarray}
   \label{main:1:e:eq:2}
   \dot {\vect E} & = & - \kappa_{\mathrm{out}}
   M\vect  E
   - i\Omega \vect E - i\kappa_{\mathrm{out}} \alpha
   \vect E + \kappa_{\mathrm{out}} (1 + i\alpha )  G (A \vect E) + \vect W, \\
    \label{main:2:D:eq:2}
    \dot D & = & -\gamma_1\left\{ D - j +
      \mathrm{Im}\left[(i - \alpha )
        \vect  E^\ast  {\cal L} (A \vect E)\right]\right\}, \\
    \label{main:3:d:eq:2} \dot d &  = & - \gamma _s d -
    \gamma_1 \mathrm{Re}\left[(i - \alpha ) \vect E^\ast
      {\cal L} (A \vect E')\right].
 \end{eqnarray}
\end{widetext}

 Here $\alpha$ is the line width enhancement factor, $j$ is the
 normalized current density, $\kappa_{\mathrm{out}}$ is the
 outcoupling losses (see below),$\gamma_1=1/T_1$ and $\gamma_s$ are
 the decay rates of $D$ and $d$, $\vect E' = \{E_y, -E_x\}$ and $ A =
 \bigl(\begin{smallmatrix} D & i d \\ -i d & D\end{smallmatrix}
 \bigr)$. The linear operators $ M(\vect k_\bot)$, $ \Omega(\vect
 k_\bot)$, $ G(\vect k_\bot)$, $ {\cal L}(\vect k_\bot)$ describe the
 losses, diffraction and gain in the cavity, and are $2 \times 2$
 $\vect k_\bot$-dependent matrices acting on the vector field $\vect
 E(\vect k_\bot)$.

 In the present paper we assume that $ \Omega(\vect k_\bot) =
 -k_\bot^2 v/k_0 \mathbb{1} + \left( s_1(\vect k_\bot)+s_2(\vect
   k_\bot)\right)/\tau$ is a matrix describing the dispersion relation
 given by the cavity resonance condition for s- and p- waves of DBR.
 Here $v$ is the speed of light in the cavity, $k_0$ is the
 longitudinal part of the wavevector, $\tau$ is the cavity round-trip
 time, $\mathbb{1}=\diag{(1,1)}$ ($\diag{(\cdot,\cdot)}$ is here a
 diagonal matrix with corresponding entities on the diagonals), and
 $s_1,\,s_2$ are the matrices, describing diffraction of the light in
 the DBRs. In s-p-representation they can be written as
 $s_i=\diag{(s_{si},s_{pi})}$, where $i=1,2$ and $s_{si}$,$s_{pi}$
 are the phase shift for s- and p-waves.

 Using these assumptions, $M(\vect k_\bot)$ and $G(\vect k_\bot)$ can
 be written in terms of propagation matrices $F_1(\vect k_\bot)=(\rho
 \tilde \Gamma)^{1/2} R_1(\vect k_\bot) \exp(-i s_1(\vect k_\bot))$,
 $F_2(\vect k_\bot)=(\rho \tilde \Gamma)^{1/2} R_2(\vect k_\bot)
 \exp(-i s_2(\vect k_\bot))$, $F=F_1F_2$: $ M(\vect k_\bot)= \left( 1-
   F(\vect k_\bot)\right)/M_0, G(\vect k_\bot) =\left( 1+ F_1(\vect
   k_\bot) +F_2(\vect k_\bot) + F(\vect k_\bot)\right){\cal
   L}_\Omega/G_0$.  $M$ and $G$ are normalized by the constants $M_0$
 and $G_0$ in such a way, that $M_{11}(0)=1$, $G_{11}(0)=1$. Here
 $R(\vect k_t)$ is an operator describing the reflection from the
 Bragg mirrors, represented by matrices $R_m={R_{mij}} \, (m=1,2,
 i=1,2,
 \,j=2,2)$. $\kappa_{\mathrm{out}}=(1-|R_{111}(0)||R_{211}(0)|)/\tau=
 (1-|R_{122}(0)||R_{222}(0)|)/\tau$ is the outcoupling losses for zero
 transverse mode. $\rho$ describes  all intracavity losses in the
 whole system (which are not due to outcoupling) and $\tilde \Gamma$
 is the intracavity anisotropy matrix, which in the basis of the main
 anisotropy axes is written as
\begin{equation} \tilde \Gamma= \left(
    \begin{smallmatrix} 
      \exp(\gamma_a\tau+i \gamma_p\tau) & 0  \\
      0 & \exp(-\gamma_a\tau-i \gamma_p\tau)   \\
    \end{smallmatrix} \right)
\end{equation}
where $\gamma_a$ is the amplitude anisotropy and $\gamma_p$ is the
birefringence.  The influence of the gain contour line ${\cal L}(\vect
k_\bot)$ is given by the expression ${\cal L} = 1/\left(1 + ((\delta -
  \Omega)/\gamma)^2\right)$ where $L$ is the cavity length and
$\gamma$ is the material polarization decay rate.

The spontaneous emission is described by the term $\vect W = \sqrt{K
  \beta_{sp}\gamma_1 D } \boldsymbol \xi$ in the approximation of zero
inter-sub-band population difference. Here, $K$ is the Petermann excess
quantum noise factor and $\beta_{sp}$ is the spontaneous emission
factor. $\boldsymbol \xi$ is the Langevin noise source with zero mean and
correlation in $(x,y)$-space and circular wave basis $\boldsymbol \xi(x,y,t)
= \{\xi_+(x,y,t), \xi_-(x,y,t)\}$:
\begin{gather}
  \langle \xi_\pm(x,y,t)
  \xi_\pm(x',y',t') \rangle = 2 \delta(t-t')\delta(x-x')\delta(y-y'),
  \, \notag \\
  \langle \xi_\pm(x,y,t) \xi_\mp(x',y',t') \rangle = 0. \label{eq:ksi}
\end{gather}
Performing the transverse Fourier transform and transforming into a
basis of linear (orthonormal) polarization with arbitrary directions
of the axes $1$ and $2$ one obtains the analogous equation for $\boldsymbol
\xi (k_x,k_y,t) = \{\xi_1(k_x,k_y,t), \xi_2(k_x,k_y,t)\}$:
\begin{equation}
  \label{eq:ksik}
  \begin{split}
    \langle \xi_{j}(k_x,k_y,t) \xi_{i}(k'_x,k'_y,t') \rangle = 2
    \delta_{ij} \delta(t-t')\times \\ \times \delta(k_x-k_x')\delta(k_y-k_y').
  \end{split}
\end{equation}
Therefore, the noise is also correlated in the $\vect
k_\bot$-representation and in an arbitrary orthogonal polarization
basis.  The noise terms in the equations for $D$ and $d$ are neglected
in the present consideration.

\subsection{The derivation of the Ginsburg-Landau equation (GLE) for
  the field}

Here we obtain the lowest-order nonlinear equations for the field
resulting from the above mentioned nonlinear equations.  We take into
account that near lasing threshold the resulting linear operator
acting on $\vect E$, $D$, $d$ has a block-diagonal form with the part
acting on $\vect E$ not being coupled to the carrier part.  The
eigenvalues stemming from the carrier related part are always strongly
negative.  In addition, the solution with $d=0$ is always
possible. Then, it is possible to adiabatically eliminate $D$ and
obtain a complex equation for $\vect E$ only.

The solution of $\dot{D}=0$ is given then by $D= j/(1+I {\cal L})$,
where $I = |\vect E|^2$. For small intensity $I$ one can write it as
$D \sim j(1-I {\cal L})$. Substituting this into the equation for the
field, we obtain:
\begin{equation}
  \label{eq:beta0}
  \begin{split}
    \dot{\vect E} = -\kappa_{\mathrm{out}} M \vect E - i \Omega \vect E -
    i\kappa_{\mathrm{out}} \alpha \vect E + \kappa_{\mathrm{out}} (1 + i\alpha )j G{\cal
      L} \vect E - \\ -\kappa_{\mathrm{out}}(1 + i\alpha )j G {\cal L}^2 I \vect E
    + \vect W.
  \end{split}
 \end{equation}

 Because the anisotropy and the intracavity losses are small compared
 to $1$, the corresponding terms can be decomposed as $\rho =
 \exp(-\kappa_{\mathrm{in}} \tau) \approx 1 - \kappa_{\mathrm{in}} \tau$,
 $\tilde \Gamma \approx 1 + \Gamma \tau$ with $\Gamma=
 \diag{(\gamma_a+i \gamma_p, -\gamma_a-i \gamma_p)}$.  Considering
 $\kappa_{\mathrm{in}} \tau$, $\kappa_{\mathrm{out}} \tau$ and $\Gamma
 \tau$ as small parameters and neglecting these terms starting from
 the first order in $G$ and from the second order in $M$, and
 introducing the matrices ${\cal G} = (1+i\alpha) j \tilde G {\cal
   L}^2$, $\Upsilon= \tilde M + i\alpha -(1+i\alpha)j \tilde G {\cal
   L}$ (where $\tilde M = M \left|_{\rho=1,\tilde \Gamma=1}\right.$,
 $\tilde G = G \left|_{\rho=1,\tilde \Gamma=1}\right.$) we obtain the
 resulting GLE (\ref{eq:egl0}).

The spontaneous emission term in the approximation of a small
intensity can be written as $\vect W(\vect k_\bot,t) =
\sqrt{K\beta_{sp}\gamma_1j} \boldsymbol \xi(\vect k_\bot,t)$. Here we
neglected the second order term in decomposition of stationary value of
$D$ into series.



\end{document}